\documentclass[aps,prl,twocolumn]{revtex4}
\usepackage{graphicx}
\usepackage{natbib}

\begin{document}
\textheight 8.8in
\textwidth 6.5in
\topmargin -.25in
\oddsidemargin -.25in
\evensidemargin 0in
\newcommand{\lsim}{\,\lower2truept
\hbox{${<\atop\hbox{\raise4truept\hbox{$\sim$}}}$}\,}
\newcommand{\gsim}{\,\lower2truept
\hbox{${>\atop\hbox{\raise4truept\hbox{$\sim$}}}$}\,}
\title{{\large {\bf Instability of Chaplygin gas trajectories in 
unified dark matter models}}}
\author{F. Perrotta$^{1,2}$, S. Matarrese$^{3,4}$, M. Torki$^{5}$} 
\affiliation{
$^1$ SISSA/ISAS, Via Beirut 4, 34014 Trieste, Italy; \\
$^2$ INFN-Sezione di Trieste, via Valerio 2, 34127 Trieste, Italy;  \\
$^3$ Dipartimento di Fisica `G. Galilei', Universita di Padova, via 
Marzolo 8, 35131 Padova, Italy; \\
$^4$ INFN-Sezione di Padova, via Marzolo 8,  35131 Padova, Italy; \\
$^5$ Dipartimento di Astronomia, Universita' di Padova,
vicolo dell'Osservatorio 2, 35122, Padova, Italy}

\newcommand{\etal}{{et~al.}}
\def\aa{{\sl Astron.\ \&\ Astrophys.}}
\def\aas{{ Astron. \& Astrophys.\ Suppl. \ }}
\def\apj{{Astrophys. \ J. \ }}
\def\apjl{{\sl Astrophys.\ J.\ Lett.}}
\def\apjs{{\sl Astrophys.\ J.\ Supp.}}
\def\araa{{\sl Ann.\ Rev.\ Astron.\ Astrophys.}}
\def\astroph#1{{\tt astro-ph/#1}}
\def\ieeespl{{\sl IEEE\ Signal\ Processing\ Lett.}}
\def\jcap{{\sl J.\ Cosm.\ Astroparticle\ Phys.}}
\def\mnras{{\sl MNRAS}}
\def\na{{\sl New\ Astron.}}
\def\np{{ Nucl.\ Phys.}}
\def\plb{{ Phys.\ Lett.\ B \ }}
\def\procieee{{\sl Proc.\ IEEE}}
\def\pr{{\sl Phys.\ Rep.}}
\def\prd{{ Phys.\ Rev. \ D \ }}
\def\prl{{ Phys.\ Rev.\ Lett.\ }}
\def\ptp{{Prog.\ Theor.\ Phys.}}
\def\rmp{{\sl Rev.\ Mod.\ Phys.}}

\begin{abstract}  
In the past few years, the Chaplygin gas (CG) has been considered as an 
appealing candidate for unifying dark matter and dark energy into a 
single substance. This picture had to face several problems when trying to
fit its predictions with cosmological observations. We point out another  
potential problem of this model, showing that, when the CG is described 
through a classical scalar field, the corresponding trajectories are strongly 
unstable. This implies that an extreme fine tuning must apply to the 
initial values of the field in order to end up today with values 
allowed by observational data.  
\end{abstract}

\maketitle

Since the observations of distant type Ia Supernovae pointed towards an 
accelerated expansion of the Universe \cite{Perl,Riess}, a great effort 
has been done to provide a reasonable, physically motivated responsible  
for this speeding up. Apparently, a dark energy component adds 
up to the dark matter, coming to dominate the total energy budget
at very recent times. The remarkable new feature of dark energy 
is that it appears to violate the strong energy condition 
\cite{Bahcall}; understanding the nature of dark energy is probably one 
of the most important open problems in modern physics.
Besides the possibility of a quintessence scalar field \cite{Quintessence},  
modifications of gravity \cite{EQ}, or an uncanceled cosmological 
constant \cite{Lambda}, an interesting alternative class of dark 
energy models  is that of the so-called Chaplygin gas 
\cite{Kamen}-\cite{Chimento2}. 
In its simplest formulation, the Chaplygin gas is a  perfect fluid 
with equation of state $p=-{A / \rho}$, $A$ being a positive constant 
with the dimensions of an energy 
squared. A generalized form of the Chaplygin gas, containing 
an additional free parameter, has also been studied in detail 
(see, e.g., \cite{Bento,Bento2}). \\
While first introduced in a hydrodynamics context \cite{Chap_original}, 
the Chaplygin gas has recently raised growing interest in particle 
physics, thanks to its connection with string theory   
\cite{Bilic,Chap_string,Sen,Gibbons,Padmanabhan,Bagla,Gorini3} and 
because it is the only fluid known to admit a supersymmetric 
generalization \cite{Chap_susy}. \\
From a cosmological point of view, the striking 
feature of the Chaplygin Gas (hereafter, CG) is that it allows for a 
very elegant unification of dark energy (DE) and dark matter (DM), since its 
equation of state interpolates between a dust dominated 
phase at early times and a de Sitter phase at late times. \\
Despite the appeal of these intriguing features, several fatal 
drawbacks turned out to affect CG models. 
There are, actually, two different approaches to the cosmological role 
of the CG: one can consider it as a unified dark matter component, so that 
$\Omega_{DM}=0$ and the CG plays both the roles of dark matter and of 
a cosmological constant, at different epochs (see, e.g., Ref.~\cite{Sandvik}); 
or one can look at the CG 
as a Dark Energy candidate, which adds to the standard DM component, as 
well as to baryons and radiation \cite{Bean,Amendola}. In both cases, 
the CG is unable to match several observational data. \\
In  Ref.~\cite{Sandvik} it was pointed out that perturbations in the CG  
should affect the formation of structures, producing oscillations or  
exponential blowups in the matter power-spectrum which are inconsistent 
with observations; the analysis ruled out the CG as a unified dark 
matter candidate ($\Omega_{DM}=0$) at $99.99 \%$ level. \\
The impact on cosmology has also been investigated in the context of 
Supernovae data \cite{Fabris,Avelino,Beca,Makler,Silva,Bertolami}, 
showing initially a certain degree of consistency with data for a class 
of CG models;  Cosmic Microwave Background (CMB) measurements, though,  
provided stronger constraints to CG cosmologies 
\cite{Carturan,Bean,Amendola}.
While Ref.~\cite{Bean} found that the joint analysis of type Ia Supernovae 
and CMB data only allows for a CG gas which is today indistinguishable from a 
cosmological constant, the  analysis of Ref.~\cite{Amendola} rules out the 
CG as a Dark Energy candidate at $99.99\%$ confidence level, and 
also indicates that the CG as a unified model of dark matter and dark energy 
is strongly disfavoured by the latest CMB data. 

In this paper, we point out another possible problem related to the 
dynamical behavior of the CG gas. We focus on the unified dark 
matter cosmological model and we analyze the trajectories of 
the classical scalar field which is meant to correspond to the CG  
hydrodynamical representation, searching for attractor solutions. 
For a flat universe, the parameters of the CG can be easily related to 
the expansion rate today, $H_0$, and to the present cosmic equation of state, 
$w_0$.
The latter quantity is constrained by the recent estimates of the 
present deceleration parameter $q_0$ \cite{Riess2}, since 
$w_0=(2q_0-1)/3$ ; we thus require $w_0 \sim -0.7$ today. 
We find that, in order to finish up at the present epoch with such an 
equation of state $w_0$ (different than $-1$) and  a plausible value of the 
expansion rate, a strong fine-tuning must be applied on the initial conditions 
of the scalar field. In 
other words, the trajectories which reproduce the observed parameters 
are strongly unstable under variations of the initial conditions on the 
field. The only attractor of the dynamical system under study is 
indeed the de Sitter Universe, to which the model converges at late 
times.  
Our conclusions are complementary to the perturbation analysis of 
\cite{Sandvik}, showing that, already at the background level, the 
unified dark matter model shows serious instabilities. 

We will undertake our analysis in a flat Friedmann-Robertson-Walker (FRW) 
cosmology; 
since we are mostly interested in the dynamics at recent times, radiation 
is not included in our treatment. 
From the energy conservation and the CG equation of state, it follows 
that 
\begin{equation}
\label{rhoChap}
\rho_{Chap}=\sqrt{A+{B \over a^{6}}} \ \ , \ \ 
p_{Chap}={-A \over \sqrt{A+{B \over a^{6}}} }
\end{equation}
$a$ being the scale factor and $B$ an integration constant;
the ratio $B/A$ characterizes the transition between the matter-like 
behavior and the cosmological-constant. \\  
In the unified dark matter model, 
the expansion rate and the cosmic equation of state today are, 
respectively, 
$ H_0^2=\sqrt{A+B}$
and
$w_0=-(1+B/A)^{-1}$ . 
We see that the constants $A$ and $B$ can be determined once $H_0$ and 
$w_0$ are known. 
Our aim is to provide a description of the Chaplygin gas 
through a minimally-coupled, classical scalar field.   
In \cite{Kamen},\cite{Gorini} this description is simply built from 
the analogy of the energy density and pressure of the CG in 
Eq.~(\ref{rhoChap}) with the analogous quantitites for a scalar field $\phi$ 
evolving in a potential $V(\phi)$, i.e. 
\begin{equation}
\rho_{\phi}= {\dot{\phi}^2 / 2 +V(\phi)} 
\ \  , \ \ 
p_{\phi}= {\dot{\phi}^2 / 2 -V(\phi)} \ \ , 
\end{equation} 
where a dot denotes differentiation w.r.t. the cosmic time. \\ 
Namely, by requiring that $\rho_{\phi}=\rho_{Chap}$ and 
$p_{\phi}=p_{Chap}$, we can relate the 
cosmic scale factor to the kinetic and potential energy a scalar field 
should have in order to reproduce the Chaplygin fluid dynamics: 
\begin{equation}
\label{phidotvsa}
\dot{\phi}^2= {B \over {a^6 \sqrt{A+{B \over a^6}}}} \ \ , \ \ 
V(\phi(a))= {2 a^6 A + B \over {2 a^6 \sqrt{A+{B \over a^6}}} } \ \ .
\end{equation}

By making use of the Friedmann equation  and of Eq.~(\ref{phidotvsa}), 
one can infer $d \phi / da$, which can be solved to give $a(\phi)$; 
The authors of Ref.~\cite{Kamen} give 
\begin{equation}
\label{avsphi}
a^6={4B {\rm{exp}}(6 \phi) \over A (1-{\rm{exp}}(6 \phi))^2} \ \ .
\end{equation}
There is a crucial assumption in Eq.~(\ref{avsphi}): the integration 
constant has been choosen so that the value of the scalar field at some 
(arbitrary) time $t_i$ is fixed; in particular, setting the initial 
conditions at $a_i$, one should have 
\begin{equation}
\label{phiinit}
\phi_i={1\over 6} {\rm ln } \left| 
{ \sqrt{ {A \over B }a_i^6+1}  -1 \over  \sqrt{ {A \over B }a_i^6+1} +1  
}
\right|  \ \ ;
\end{equation}
The initial ``velocity'' $\dot{\phi}_i$ is fixed too, by 
Eq.~(\ref{phidotvsa}). Substituting $a(\phi)$ in $V(a)$, one has 
\begin{equation}
\label{Vvsphi}
V(\phi)= {1 \over 2} \sqrt{A} \left( {\rm{cosh}} 3 \phi + {1 \over 
{\rm{cosh}} 3 \phi } \right) \ \ .
\end{equation}
As already noticed in Ref.~\cite{Gorini2}, a scalar field evolving in such a 
potential will produce a cosmological evolution coinciding with that of 
the Chaplygin gas only if the initial conditions are appropriately 
handled to satisfy the relation $\dot{\phi}_i^4=4(V^2(\phi_i)-A)$; 
indeed, the potential (\ref{Vvsphi})  has been built under this assumption. 
\\  
The evolution of a minimally-coupled scalar field follows the 
Klein-Gordon equation of motion:
\begin{equation}
\label{KG}
\ddot{\phi}=-3 H \dot{\phi}-V_{,\phi}   \ \ .
\end{equation}
As we said, in order for the solution of this equation to reproduce the  
energy density and pressure (\ref{rhoChap}), one has to 
exactly fix the initial values of $\phi,\dot{\phi}$.
This is quite  intuitive, and it can also be easily understood by looking at 
the shape of the potential (\ref{Vvsphi}), which is extremely flat for 
small values of  the field, increasing exponentially elsewhere; as an 
example, a scalar field initially set on the minimum of this potential 
with vanishing velocity, would exhibit no dynamics at all, and it would be 
indistinguishable from a cosmological constant at any cosmic epoch. 
Of course, this is not what we are expecting from a 
plausible scalar-field representation of the Chaplygin fluid; first, 
a unified model of dark matter needs at least an epoch when the 
equation of state 
$w_{\phi}\equiv p_{\phi}/\rho_{\phi}$ 
is zero, so that the energy density drops like a matter component. 
This phase should 
end up in a de Sitter phase, when the energy density of the scalar 
field is constant (as for $a >> 1$ in (\ref{rhoChap}). Finally, since 
the cosmic equation of state today is kinematically related to the  
deceleration parameter $q \equiv \ddot a/aH^2$, the most recent 
measurements give $w_0\sim 0.7$ \cite{Riess2}. It follows that 
the intermediate stage of the CG evolution should be characterized by  
a decreasing equation of state, shifting smoothly from $0$ to $-1$.
The point we focus on here is how much fine-tuning, if any, is 
required to do that. In other words, is there any freedom in setting 
the initial values of the field if one has to reproduce the scalings in 
Eq.~(\ref{rhoChap})? The question may be turned into the phase-space 
language, by asking whether the Chaplygin fluid trajectory is an 
attractor for the system (\ref{KG}). 
As we said, we can always provide a scalar-field description of the 
Chaplygin gas, by appropriately setting its initial conditions and 
building up the corresponding potential. Such a field will obey the 
Klein-Gordon equation, and, in addition, 
$\rho_{\phi},p_{\phi}$ will scale as in Eq.~(\ref{rhoChap}) 
at any time. 
The scalar field model obtained with this particular choice of 
initial conditions 
would perfectly reproduce the dynamics of the Chaplygin gas. We will 
refer to $\phi_{Chap}(t)$ as the solution of Eq.~(\ref{KG}) with 
potential (\ref{Vvsphi}), which satisfies the required initial 
conditions.   

As for the late time behavior of (\ref{KG}), it is easily seen 
that, whatever initial conditions we choose, the system will end up in 
its global attractor, the de Sitter stable node (see \cite{SZY}) with 
$\phi,\dot{\phi}=0$, $H=\sqrt{A}$ (note that the de Sitter node is the 
only critical point of the dynamical system under investigation). In 
particular, this is true for the solution $\phi_{Chap}(t)$.
However, we are interested in finding whether the trajectory 
$\phi_{Chap}$ itself is an attractor for the problem on hand, 
in the sense that many possible solutions of the system 
converge to a common evolutionary track 
represented by the $\phi_{Chap}$ trajectory.   
We find that this is not the case; while $\phi_{Chap}$ requires
fine-tuned initial values to reproduce the Chaplygin gas cosmology,  
slightly different initial conditions on $\phi_{Chap}$ may result in 
completely different trajectories, which may even be unable to 
reproduce the matter behavior at all. The general solutions of this 
dynamical system turn out to behave much differently than 
the ``tracking solutions'' of 
\cite{gamma}. Following the notation of \cite{gamma}, we may anticipate 
our result, saying that the system does not admit tracking solutions 
because one of the two necessary 
conditions on $\Gamma \equiv {V V_{,\phi \phi} / V_{,\phi}^2}$ is 
violated, namely $\Gamma=1$ for almost any plausible initial field 
value. One should be careful, however, in using the formalism of 
\cite{gamma} in this case, in that the Chaplygin gas is supposed to be 
the only component of the cosmic fluid, and there is no other 
background component which may drive its trajectories (having
neglected the role of radiation in our analysis). For this reason 
we found it useful to directly check for the stability of the 
$\phi_{Chap}$ solutions with a numerical integration of the system.         

\begin{figure}
\includegraphics[width=3.5in, height=2.8in]{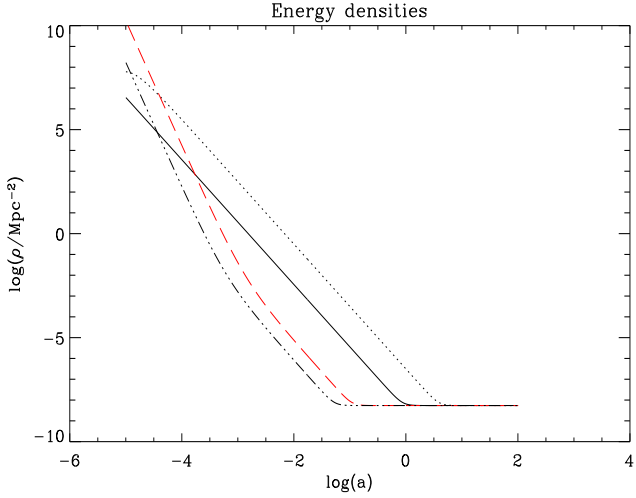}
\caption{Energy densities for the $\phi_{Chap}$ solution and for 
different initial conditions, corresponding to the 
cases a),b),c) described in the text. }
\label{Figenergy}
\end{figure}

\begin{figure}
\includegraphics[width=3.5in, height=2.8in]{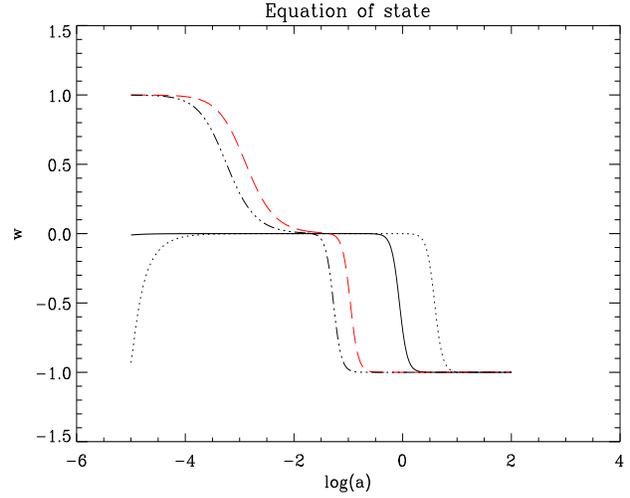}
\caption{Equation of state for the $\phi_{Chap}$ solution and for
different initial conditions, corresponding to the
cases a), b), c) described in the text.}
\label{Figeos}
\end{figure}

Our plots refer to four different choices of initial conditions, 
set at redshift $z \sim 10^{5}$.	
The solid lines refer to the $\phi_{Chap}$ trajectories. 
The present values of $H_0$ and $w_0$ unambiguously 
determine $A$ and $B$ for a flat universe and the 
present values of $\phi_{Chap}, \dot{\phi}_{Chap}$. We set $h=0.7$ 
and  $w_0=-0.7$. 
Together with the $\phi_{Chap}$ trajectories,
Figs.~\ref{Figenergy} and \ref{Figeos} show, respectively, 
the energy density of the scalar 
field and the corresponding equation of state, when the field starts 
with the following initial conditions: \\
a) $\phi_i=\phi_{Chap,i}\times 1.1, \ 
\dot{\phi}_{i}=\dot{\phi}_{Chap,i}$ (dotted lines); \\
b) $\phi_i=\phi_{Chap,i}, \ \dot{\phi}_{i}=\dot{\phi}_{Chap,i} \times 
10^{2}$ (dashed lines); \\
c) $\phi_i=\phi_{Chap,i}\times 0.9, \ 
\dot{\phi}_{i}=\dot{\phi}_{Chap,i}\times 10 $  
(dot-dashed lines).  \\
Even a slight modification of the initial values with 
respect to $\phi_{Chap}$ can make the 
equation of state today completely different from the required 
value, resulting in trajectories which never join the $\phi_{Chap}$ 
curves, nor reproduce the CG behavior at any epoch. 
Furthermore, the conditions a), b), c) finish up at present with 
unacceptable values of the Hubble expansion rate.  
Starting with a bigger amount of kinetic energy, as in b) and c),   
an initial kination precedes the phase in which the field behaves like 
matter. 
In each of the plotted cases, there is a very sharp transition between 
the matter-like scaling ($w=0$) and the $\Lambda$-like one ($w=-1$): 
because of this sharpness, values of the equation of state different than 
$0,-1$ turn out to be unnatural. This behavior is very different than 
in tracking quintessence solutions, where the only parameters which need 
to be adjusted are the amplitude of the potential and the current equation 
of state: in that case, once these present values are fixed, there is no 
dependence on the initial values of the quintessence field.  

Concluding, we have shown that, in order to have a unified dark matter 
model consistent with the most recent estimates of the cosmic equation 
of state, one is forced to introduce an extreme fine tuning on 
the initial values of the scalar field, because the problem does not 
admit tracking solutions; the de Sitter universe is the only 
global attractor of the system. \\
Although the potential (\ref{Vvsphi}) requires ad hoc initial 
conditions on $\phi$, which allowed us to define $\phi_{Chap}$, this is 
not a unique representation of the Chaplygin fluid. 
One may start with different values of $\phi_i,\dot{\phi}_i$ and, 
by a similar approach, build a potential $V(\phi)$ (generally, different 
from \ref{Vvsphi}) for each set of initial conditions, so to reproduce the 
behavior \ref{rhoChap}. 
In this way, there would be an infinite class of potentials, mapping the 
behavior of the Chaplygin fluid. However, our conclusions apply to any 
on this potentials, due to the peculiar sharpness of the matter-$\Lambda$ 
transition of the CG fluid. \\
Therefore, the CG as a unified dark matter candidate, seems to be 
strongly disfavored from the point of view of its dynamics. \\

{\it{Acknowledgments}} \\
{F.P. wishes to thank Carlo Baccigalupi and Martin Makler for the 
useful comments.} 

\bibliographystyle{}

\end{document}